\documentclass[amsmath,amssymb,prb,superscriptaddress,twocolumn,showpacs]{revtex4-1}
\usepackage{graphicx}
\usepackage{dcolumn}
\usepackage[colorlinks=true,linkcolor=blue,citecolor=blue,urlcolor=blue]{hyperref}
\usepackage{stmaryrd}
\usepackage{esvect}
\usepackage{multirow}
\usepackage{upgreek}
\allowdisplaybreaks

\renewcommand{\vec}[1]{\mathbf{#1}}
\newcommand\dd{\text{d}}

\usepackage{siunitx}
\DeclareSIUnit\mub{\mu_\text{B}}
\DeclareSIUnit\rydberg{\text{Ry}}
\usepackage{mathtools}

\DeclarePairedDelimiterX\braket[1]{\langle}{\rangle}{#1}

\usepackage[usenames,dvipsnames]{xcolor}

\begin{document}

\title{Spin, atomic and inter-atomic orbital magnetism induced by 3\textit{d} nanostructures deposited on transition metal surfaces}

\author{Sascha Brinker}\email{s.brinker@fz-juelich.de}
\affiliation{Peter Gr\"{u}nberg Institut and Institute for Advanced Simulation, Forschungszentrum J\"{u}lich \& JARA, 52425 J\"{u}lich, Germany}
\affiliation{Department of Physics, RWTH Aachen University, 52056 Aachen, Germany}
\author{Manuel dos Santos Dias}\email{m.dos.santos.dias@fz-juelich.de}
\affiliation{Peter Gr\"{u}nberg Institut and Institute for Advanced Simulation, Forschungszentrum J\"{u}lich \& JARA, 52425 J\"{u}lich, Germany}
\author{Samir Lounis}\email{s.lounis@fz-juelich.de}
\affiliation{Peter Gr\"{u}nberg Institut and Institute for Advanced Simulation, Forschungszentrum J\"{u}lich \& JARA, 52425 J\"{u}lich, Germany}

\date{\today}

\begin{abstract}
\noindent
We present a first-principles study of the surface magnetism induced by Cr, Mn, Fe and Co adatoms on the (111) surfaces of Rh, Pd, Ag, Ir, Pt and Au.
We first describe how the different contributions to the surface magnetism enter the magnetic stray field, with special attention paid to the induced orbital moments.
Then we present results for the spin and orbital magnetic moments of the adatoms, and for the induced surface spin and orbital magnetic moments, the latter being further divided into atomic and inter-atomic contributions.
We investigate how the surface magnetism is determined by the chemical nature of the elements involved, such as the filling of the magnetic d-orbitals of the adatoms and the properties of the itinerant electrons at the surface (whether they are sp- or d-like, and whether the spin-orbit interaction is relevant), and how it is modified if the magnetic adatoms are brought together to form a cluster, with Cr, Mn, Fe, and Co trimers on Pt(111) as an example.
We also explore the impact of computational approximations, such as the distance between the adatom and the Pt(111) surface, or confinement effects due to the finite thickness of the slab used to model it.
Our discussion of the magnetic stray field generated by a single adatom and its environment suggests a possible way of disentangling the induced surface magnetism from the adatom one, which could be feasible with scanning NV-center microscopy.
\end{abstract}

\maketitle

\section{Introduction}
The properties of magnetic adatoms and small clusters deposited on non-magnetic metallic surfaces are of great interest for fundamental physics and for potential technological applications.
The same magnetic adatom can behave very differently when placed on different surfaces, which highlights the importance of its interaction with the surface electrons, and how they react to its presence.
This is typified by Friedel oscillations~\cite{friedel_metallic_1958} of the charge and spin densities around the adatom~\cite{Crommie1993}, which can mediate long-ranged Ruderman-Kittel-Kasuya-Yosida interactions~\cite{Meier2008,Zhou2010,Hermenau2019} between magnetic adatoms or other nanostructures.
If the surface possesses strong spin-orbit coupling (SOC), it can also enable large magnetic anisotropy energies~\cite{gambardella_giant_2003,Blonski2010,Donati2013,Singha2017}, Dzyaloshinskii-Moriya interactions~\cite{Fert1980,Menzel2012,Khajetoorians2016}, or even chiral biquadratic interactions~\cite{Brinker2019}.
Knowledge of these interactions is necessary to understand what kind of magnetic states are stabilized and how they can be manipulated, if one envisions their usage as elementary bits for information storage and processing~\cite{Khajetoorians2011,Loth2012,Khajetoorians2013,hermenau_gateway_2017,Hermenau2019}.

The magnetic properties of such surface-supported nanostructures are experimentally difficult to probe.
X-ray magnetic circular dichroism provides element-specific information on the spin and orbital magnetic moments through sum rule analysis and on the magnetic anisotropy energy~\cite{gambardella_giant_2003,Blonski2010,Donati2013,Singha2017}, averaged over the sample.
At the atomic scale, scanning tunneling microscopy and spectroscopy has provided many insights, being able to manipulate, probe and excite even single atoms on surfaces~\cite{Wiesendanger2009}.
On the nanometer scale, the emerging technique of scanning NV-center microscopy can be used to detect the magnetic stray fields in the order of $\SI{10}{\nano\tesla}$ generated by a single atomic spin~\cite{Grinolds2013}.
Theoretical studies can be hampered by the potential need to account for large separations between interacting magnetic nanostructures, or for a polarization cloud surrounding the nanostructure that can involve thousands of atoms~\cite{oswald_giant_1986,zeller_large-scale_1993}.

Central to the understanding of all these phenomena is the magnetism induced on the surface by the presence of magnetic nanostructures.
This should depend critically on the nature of the surface electrons (whether they are $sp$- or $d$-orbitals) and on the strength of SOC in the surface, as this affects how strongly the magnetic nanostructure can hybridize with the surface.
The induced spin moments are expected to relate to the Stoner enhancement of the spin susceptibility of the surface, and not depend too much on SOC, as illustrated by the giant spin moments induced in Pd~\cite{low_distribution_1966,nieuwenhuys_magnetic_1975}.
The induced orbital moments are not as well understood, with SOC a key ingredient for their formation, and the induced spin moments also playing a role.
Our recent work~\cite{brinker2018} has shed light on this aspect, by proposing and applying an efficient and accurate method of computing the long-range orbital magnetism induced by magnetic adatoms on the Pt(111) surface, focusing on a newly-identified contribution, the inter-atomic orbital moment, which is generated by the net-currents flowing through the surface atoms.

In this paper, we present a comprehensive study of the surface magnetism generated by magnetic adatoms (and a trimer) deposited on different non-magnetic surfaces.
We select Cr, Mn, Fe and Co adatoms, since those have a stable magnetic state for all the different substrates (Rh, Pd, Ag, Ir, Pt, Au), which we chose to explore several aspects.
The importance of SOC for the induced magnetism is found by comparing the 4\textit{d} to the 5\textit{d} elements, the latter having a much stronger SOC.
The dependence on the spin polarizability of the surface is investigated by comparing Rh with Pd and Ir with Pt, respectively, as Pd and Pt are known to have a very large spin polarizability~\cite{low_distribution_1966,nieuwenhuys_magnetic_1975,oswald_giant_1986,zeller_large-scale_1993,herrmannsdorfer_magnetic_1996,khajetoorians_spin_2013,bouhassoune_rkky-like_2016-2}. 
In addition, Ag(111) and Au(111) host free-electron-like surface states (which are Rashba-split in the case of Au)~\cite{LaShell1996,Nicolay2001}, and can therefore be compared to other works, which modelled ground-state currents induced in a Rashba electron gas due to the presence of magnetic impurities~\cite{Bouaziz_2018}.
We also address how the surface magnetism depends on the size of the magnetic nanostructure, using a Co trimer on Pt(111) as example, and whether the induced magnetic moments can be detected and mapped through the magnetic stray fields that they generate.

\section{Theoretical framework}
	\subsection{Ground state charge currents}
	We first briefly recap the concept of ground-state charge currents, already presented in Ref.~\onlinecite{brinker2018}.
	Consider a non-relativistic single-particle Hamiltonian of the form
	\begin{align}
		\mathcal{H} = \frac{1}{2m_e} \left(\vec{p} + e \vec{A}(\vec{r})\right)^2 + V(\vec{r}) + \mu_\mathrm{B} \boldsymbol{\upsigma} \cdot \vec{B}(\vec{r}) \quad ,
	\end{align}
	where $m_e$ is the electron mass, $e$ is the elementary electric charge, $V(\vec{r})$ is an electrostatic potential, $\mu_\mathrm{B} = \frac{e\hbar}{2m_e}$ is the Bohr magneton, $\vec{B}(\vec{r})$ is a magnetic field and $\vec{A}(\vec{r})$ is the corresponding vector potential.
	The Heisenberg equation of motion applied to the charge density $\rho(\vec{r},t)$, $\mathrm{i}\hbar\,\frac{\partial\rho}{\partial t} = [\mathcal{H},\rho]$,  yields the charge continuity equation, which naturally defines the ground-state charge current in terms of a para- and diamagnetic contribution\cite{maekawa_spin_2012},
	\begin{align}
	\vec{j}(\vec{r}) = \vec{j}_\text{para}(\vec{r}) +  \vec{j}_\text{dia}(\vec{r}) \quad .
	\end{align}
	In the following we will drop the diamagnetic part, since we do not consider external magnetic fields.
	The paramagnetic contribution is given by
	\begin{align}
		\vec{j}_\text{para} (\vec{r}) = -\mathrm{i}\,\mu_\text{B} \left[\Psi^\dagger(\vec{r}) \left( \vec{\nabla} \Psi(\vec{r}) \right)- \left(\vec{\nabla}\Psi^\dagger(\vec{r}) \right) \Psi(\vec{r}) \right] ,
	\end{align}
	where $\Psi(\vec{r})$ is the single-particle wave function.
	As we will employ a Green function formalism, we rewrite this expression as
\begin{align}
    	\vec{j}(\vec{r}) &= -\mathrm{i}\,\mu_\mathrm{B} \lim_{\vec{r}' \rightarrow \vec{r}} \left( \vec{\nabla} - \vec{\nabla}' \right) \text{Tr} \ \rho(\vec{r},\vec{r}') \quad , \label{current_density_matrix}
\end{align}
where
\begin{equation}\label{density_matrix}
  \rho(\vec{r},\vec{r}') = \frac{1}{2\pi\mathrm{i}}\!\int \!\dd E \ f(E) \left(G^\dagger(\vec{r}',\vec{r};E) - G(\vec{r},\vec{r}';E)\right) 
\end{equation} 
is the density matrix (a $2\times2$ matrix in spin space, $\text{Tr}$ is its trace), $f(E)$ is the Fermi-Dirac distribution function and $G(\vec{r},\vec{r}';E)$ is the retarded single-particle Green function (naturally also a $2\times2$ matrix in spin space).
	For a relativistic Hamiltonian, there is an additional contribution to the ground-state charge current due to spin-orbit coupling, which we disregard since it was found to be very small\cite{Brinker2016}.
	However, the lifting of the orbital degeneracy by spin-orbit coupling is a crucial prerequisite for the existence of finite ground-state currents.
	
	\subsection{Orbital magnetic moments}
	The classical picture of a current loop giving rise to a magnetic moment, familiar from Maxwell's laws, can also be applied to the quantum-mechanical ground-state currents.
	For systems such as the ones we consider, where the ground-state charge current density $\vec{j}(\vec{r})$ is spatially confined, it directly defines the orbital magnetic moment,
	\begin{align}
		\vec{m}_\text{o} = \mu_\text{B} \braket{\vec{L}} = \frac{1}{2} \int_\mathcal{V} \dd \vec{r}  \ \vec{r} \times \vec{j}(\vec{r}) \quad . \label{OMM_from_Maxwell}
	\end{align}
	To gain more insight and to connect with the computational method that will be employed, we divide space into cells $\mathcal{V}_i$ each containing one atom $i$.
	This leads to a separation of the orbital magnetic moment into two parts,
	\begin{align} \label{morb_loc_non_loc}
  \vec{m}_\text{o} &= \sum_i \frac{1}{2} \left( \vec{R}_i \times \vec{j}_i^\text{net}
  + \int_{\mathcal{V}_i}\hspace{-0.5em}\dd \vec{r}\;\big( \vec{r}- \vec{R}_i \big)\! \times \vec{j}(\vec{r}) \right) \nonumber \\
  &= \sum_i \left( \vec{m}_{\text{o},i}^\text{ia} + \vec{m}_{\text{o},i}^\text{a} \right)
  = \vec{m}_{\text{o}}^\text{ia} + \vec{m}_{\text{o}}^\text{a} \quad .
	\end{align}
	The atomic contribution to the orbital moment $m_\text{o}^\text{a}$ captures the swirling of the electrons around each nuclei, whereas the inter-atomic contribution to the orbital moment $m_\text{o}^\text{ia}$ describes the net currents, $\vec{j}_i^\text{net} = \int_{\mathcal{V}_i} \dd \vec{r} \ \vec{j}(\vec{r})$, flowing through a given atom.
	The atomic orbital moment is equivalent to a direct evaluation of the atomic angular momentum as shown in eq.~\eqref{OMM_from_Maxwell}.
	However, knowledge of the ground-state charge currents is crucial to obtain the inter-atomic orbital moment.
	Ref.~\onlinecite{brinker2018} reported on the inter-atomic orbital moment created by single adatoms deposited on the Pt(111) surface.
	
	\subsection{Current-induced magnetic fields}
	\label{sec:current_induced_fields}
	A possible way of measuring the microscopic distribution of the magnetic moments is by detecting the magnetic stray fields they generate.
	Classical electrodynamics \cite{jackson_classical_1999} predicts that an assembly of $N$ magnetic dipoles creates a magnetic field at some point $\vec{r}$ which is given by
\begin{align}
	\vec{B}(\vec{r}) = \frac{\mu_0}{4 \pi} \sum_i \left[ 
	\frac{3((\vec{r}-\vec{R}_i)\cdot \vec{m}_i) (\vec{r}-\vec{R}_i)}{|\vec{r}-\vec{R}_i|^5} - \frac{\vec{m}_i}{|\vec{r}-\vec{R}_i|^3}
	\right] \quad . \label{eq:stray_field_dipole_approx}
\end{align}
    Here $\mu_0$ is the vacuum permeability, and we have to sum over all $N$ magnetic atoms, with $\vec{m}_i$ the magnetic moment of the $i$-th atom located at $\vec{R}_i$. 
	This approximation holds if the distance to each atom is larger than the inter-atomic separation, $|\vec{r}-\vec{R}_i| > d$.
	
	The stray field can also be computed directly from a given ground-state charge current distribution, by performing a multipole expansion of the magnetic vector potential created by the latter,
	\begin{align}
	\vec{A}(\vec{r}) &= \frac{\mu_0}{4 \pi} \int \dd \vec{r}'\,\frac{ \vec{j}(\vec{r}')}{| \vec{r} - \vec{r}' |}  \quad .
	\end{align}
	Then the magnetic field can be obtained as
\begin{align}
\vec{B}(\vec{r}) &= \vec{\nabla} \times \vec{A}(\vec{r}) \nonumber\\
  &\approx \frac{\mu_0}{4 \pi} \sum_{i} \frac{\epsilon_{\alpha\beta\gamma} \hat{\vec{e}}_\gamma}{|\vec{r}-\vec{R}_i|^3}
 \bigg\lbrace
 \mathcal{M}_{i,\alpha} \left(r_\beta - R_{i,\beta}\right) \nonumber\\
 &\phantom{\approx} + 3\mathcal{D}_{i,\alpha\delta}\,\frac{\left(r_\delta -R_{i,\delta}\right) \left(r_\beta - R_{i,\beta}\right)}{|\vec{r}-\vec{R}_i|^2}
 - \mathcal{D}_{i,\alpha\beta}\bigg\rbrace
 \quad . \label{current_induced_bfield}
 \end{align}
Here $\epsilon_{\alpha\beta\gamma}$ is the Levi-Civita symbol, $\alpha,\beta,\gamma,\delta \in \{x,y,z\}$, $\hat{\vec{e}}_\gamma$ is the unit vector in the $\gamma$-direction, and we employ the Einstein summation convention over repeated Greek indices.
$\mathcal{M}_{i,\alpha}$ are the monopole and $\mathcal{D}_{i,\alpha\beta}$ are the dipole coefficients of the multipole expansion of the ground-state charge current around atom $i$,
\begin{align}
\mathcal{M}_{i,\alpha} &= \int_{\mathcal{V}_i } \dd \vec{r} \ j_\alpha(\vec{r}) \quad , \\
\mathcal{D}_{i,\alpha\beta} &= \int_{\mathcal{V}_i} \dd \vec{r} \ j_\alpha(\vec{r})\,r_\beta  \quad .
\end{align}
The monopole contribution is just the net current flowing through a given atomic cell, while the dipoles are related to the atomic orbital magnetic moment of each atom.
The monopoles appear in the expression for the magnetic stray field due to the multi-center nature of the multipole expansion that we employed, that is, the point chosen as the origin for the expansion is always moved to the atomic position of each atom.
If a fixed point were chosen as the origin, then there would be no monopole contribution.
In general, it still applies that $\sum_i \mathcal{M}_{i,\alpha} = 0$.

\subsection{Spin magnetic moments}
Contrary to the orbital magnetic moments, the calculation of the spin magnetic moments poses no theoretical or computational difficulty.
Atomic-like quantities can be defined by partitioning the spin magnetic moment density (which is a density of magnetic dipoles) into cells as
\begin{align}
    \vec{m}_\text{s} = \sum_i \vec{m}_{\text{s},i} = \sum_i \int_{\mathcal{V}_i}\hspace{-0.5em}\dd \vec{r}\;\vec{m}_{\text{s}}(\vec{r}) \quad .
\end{align}
The spin density itself follows from the diagonal part of the density matrix,
\begin{align}
    \vec{m}_{\text{s}}(\vec{r}) &= \mu_\mathrm{B}\,\text{Tr}\,\boldsymbol{\upsigma}\,\rho(\vec{r},\vec{r}) \quad .
\end{align}
For the purpose of calculating the magnetic stray field, the spin magnetic moments generate the standard contribution from the superposition of point magnetic dipoles, which can be straightforwardly added to Eq.~\eqref{current_induced_bfield}.

\section{Computational details}
Our first-principles investigations of surface magnetism employ two computational codes with complementary abilities.
Realistic geometries for clusters on surfaces are obtained with the Quantum Espresso (QE) package, while long-ranged induced surface magnetism can be addressed with the Korringa-Kohn-Rostoker Green function (KKR-GF) method.

Since the present KKR-GF implementation is not capable of structural relaxations, we used the plane-wave code Quantum Espresso~\cite{QE-2017} to calculate the optimized geometry of adatoms on each surface. 
The exchange-correlation effects were treated within the generalized gradient approximation using the PBEsol functional\cite{Perdew2008} with ultrasoft pseudopotentials from the pslibrary.1.0.0\cite{DalCorso2014}.
After convergence tests, we set the kinetic energy cutoff to $\SI{100}{\rydberg}$ for all the calculations.
The Monkhorst-Pack grids contained $8\times8\times8$ $k$-points for bulk calculations and $2\times2\times1$ $k$-points for supercell calculations.
The surfaces were modelled by $4\times4$ supercells containing 5 substrate layers and a vacuum region corresponding to a thickness of 6 interlayer distances, resulting in a total of 80 substrate atoms plus 1 adatom in the supercell.
These calculations were done in the collinear spin-polarized mode.

The KKR-GF calculations were performed with the potential in the atomic sphere approximation but with full charge density~\cite{papanikolaou_conceptual_2002}.
Exchange and correlation effects are treated in the local spin density approximation (LSDA) as parametrized by Vosko, Wilk and Nusair~\cite{vosko_accurate_1980}, and spin-orbit coupling is added to the scalar-relativistic approximation.
In a first step, the electronic structure of the surface without impurities is calculated.
The surfaces are modelled by a slab of 40 layers (if not explicitly mentioned) using the experimental lattice constants.
Open boundary conditions are used in the stacking direction, with the top and bottom of the slab terminated by two vacuum regions equivalent to 4 inter-layer distances.
No relaxation of the surface layer is considered, as it is known to be negligible~\cite{blonski_density-functional_2009} and was also verified by our QE calculations.
$150 \times 150$ $k$-points were used to sample the two-dimensional Brillouin zone, and the angular momentum expansions for the scattering problem are carried out up to $\ell_\text{max} = 3$.
In a second step, an embedding method is used to place each adatom in the fcc-stacking position on the surface.
The embedding region consists of a spherical cluster around each magnetic adatom, augmented with a hemisphere of substrate atoms.
The current density is efficiently evaluated by utilizing a minimal \textit{spdf} basis built out of regular scattering solutions evaluated at two or more energies, by orthogonalizing their overlap matrix~\cite{dos_santos_dias_relativistic_2015}.
More details on the calculation of the ground-state currents and details on the extraction of the inter-atomic orbital moments can be found in Ref.~\onlinecite{brinker2018}.

\section{Results}
\subsection{Structural relaxation} \label{app:structural_relaxation}

	\begin{figure}[t]
		\includegraphics[scale=1]{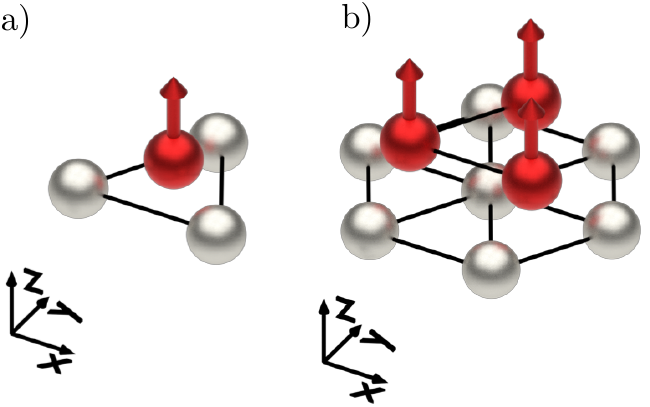}
		\caption{\label{fig:adatom_trimer_illustration} 	
		Illustration of the two considered classes of systems. a) Magnetic adatoms in the fcc-stacking position. b) Compact magnetic trimers in the fcc-top stacking position.
		The grey planes indicate the three mirror planes and the black arrows rotations of $\SI{120}{\degree}$, which are present in both system classes obeying the $C_{3v}$ symmetry.
		}
	\end{figure}
	
As explained in the previous section, the structural relaxations are obtained from QE calculations.
In a first step, the theoretical bulk lattice constant $a_0$ was calculated for all the different substrate elements --- Rh, Pd, Ag, Ir, Pt and Au, in good agreement with previous results.
In a second step, the different adatoms (Cr, Mn, Fe and Co) were deposited on a fcc(111) facet of the different substrates, in an fcc-stacking position (see Fig.~\ref{fig:adatom_trimer_illustration}a).
In this configuration, we let the adatoms as well as the first surface layer relax, whereas the other four substrate layers are kept fixed.
Since the surface atoms do not relax uniformly, but the nearest-neighbors of the adatom move typically slightly more towards the adatom than the other surface atoms, we use the mean vertical position of the first surface layer to define the mean vertical distance between the adatoms and the surface.
The results are shown in Fig.~\ref{fig:relaxations_adatoms}.
We estimated an uncertainty of approximately $\SI{0.1}{\percent}$ on the structural relaxations from the used convergence criteria for the forces.
The data for this section is collected in Table \ref{tab:relaxations} of the appendix.

\subsection{Magnetism of adatoms on (111) surfaces}
	\begin{figure}[!bt]
		\includegraphics[width=0.9\columnwidth]{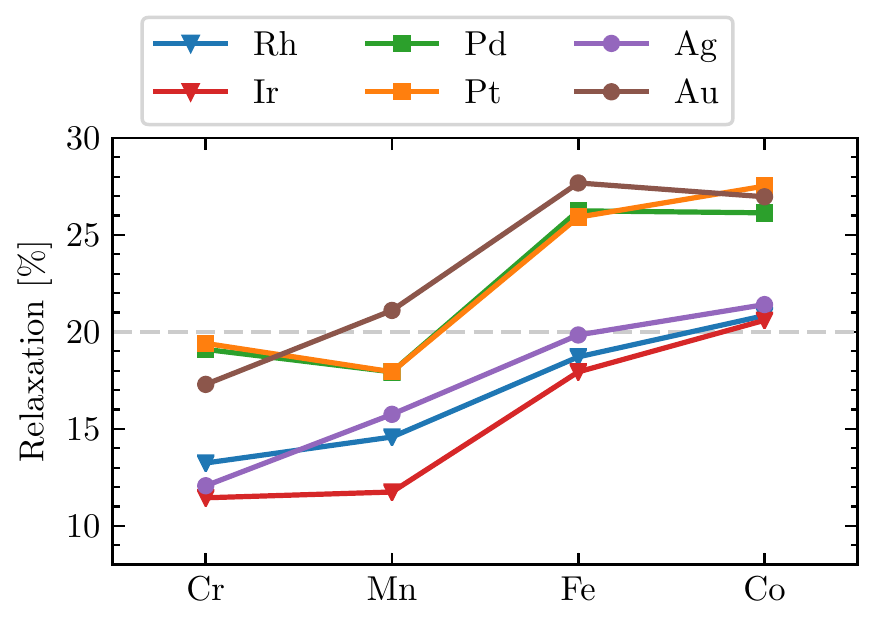}
		\caption{\label{fig:relaxations_adatoms} 	
		Relaxation of the four 3\textit{d} transition metal adatoms deposited on the fcc(111) surfaces of the six substrates.
		The vertical relaxations $\Delta$ were obtained with QE and are given as a reduction of the respective bulk interlayer distance, $d = (1 - \Delta)\,a_0/\sqrt{3}$, where $a_0$ is the bulk lattice constant.
		For the sake of comparison in the KKR-GF calculations, we chose a common relaxation value of $\SI{20}{\percent}$ (grey dashed line). 
		}
	\end{figure}

	\begin{figure}[tb]
		\includegraphics[width=\columnwidth]{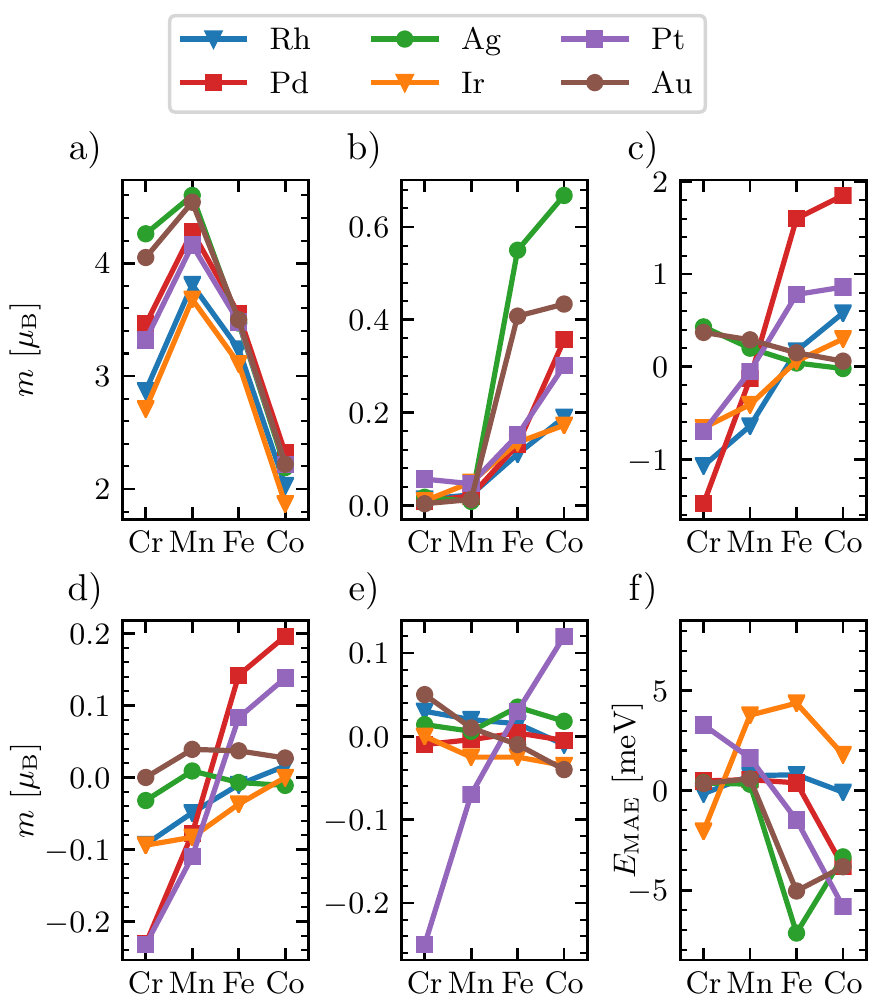}
		\caption{\label{fig:all_elements} 	
		Dependence of the magnetic contributions on the host and the adatom for Cr, Mn, Fe and Co deposited on the (111) surfaces of Rh, Pd, Ag, Ir, Pt and Au.
		a) Spin moment of the adatom. 
		b) Orbital moment of the adatom.
		c) Induced spin moments to the substrate.
		d) Induced atomic orbital moments to the substrate.
		e) Inter-atomic orbital moments in the substrate.
		All magnetic moments are in units of $\mu_\text{B}$, as indicated on the left.
		f) Magnetic anisotropy energy of the adatom.
		The band energy difference between the magnetic moments pointing in the $z$-direction and pointing in the $x$-direction was used to calculate the MAE via the magnetic force theorem.
		Positive values mean that in-plane alignment of the magnetic moment of the adatom is favored, whereas negative values favor an out-of-plane orientation.
		}
	\end{figure}
		
	\begin{figure}[tb]
		\includegraphics[width=\columnwidth]{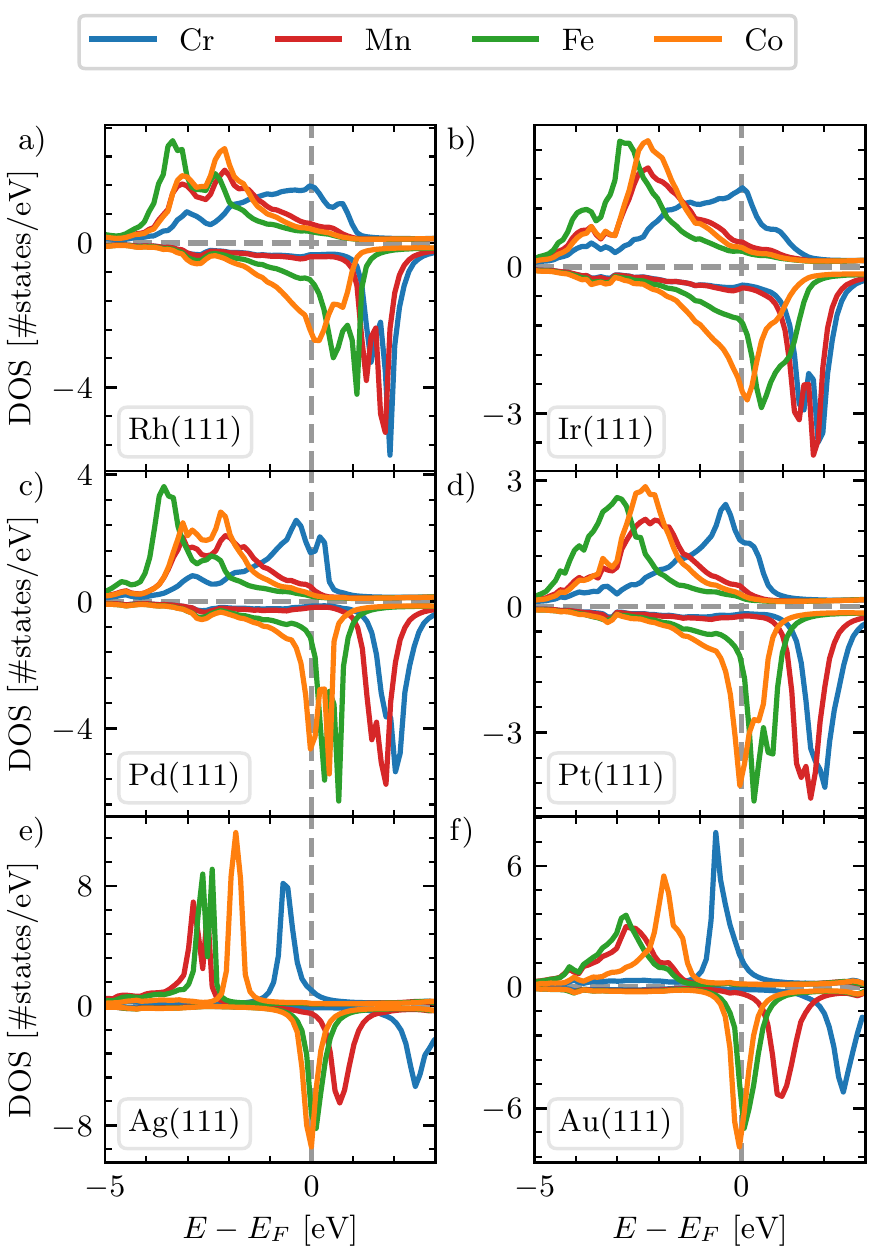}
		\caption{\label{fig:app_ldos_all_elements} 	
		Local density of states of the different adatoms deposited on the (111) facets of the selected substrates.
		Positive (negative) values correspond to the majority (minority) spin states.
		}
	\end{figure}

	Now we focus on the magnetic properties of various adatoms on several surfaces.
	For simplicity and ease of comparison, we decided to stick to a fixed vertical adatom relaxation of $\SI{20}{\percent}$ towards the surface, which is a reasonable average value for most of the substrates (see Fig.~\ref{fig:relaxations_adatoms}). 
    We will show in Sec.~\ref{sec:relaxation_dependence_Pt} that, even though the relaxation towards the surface is important for the magnetic properties, a few percent change in its value does not strongly affect the magnitude of the results and especially the trends as function of the chemical nature of the adatom.

    First, it is useful to know what is the preferred orientation of the magnetic moment of the adatoms on each surface, which is encoded in the magnetic anisotropy energy.
    This is calculated by making use of the magnetic force theorem~\cite{Liechtenstein_1987} and the frozen potential approximation.
    We perform a self-consistent calculation with an out-of-plane orientation of the adatom magnetic moment, and then a single iteration by rotating the obtained self-consistent potential to point the magnetic moment of the adatom in-plane.
    The magnetic anisotropy energy is then approximated by the difference of band energies obtained from the two calculations.
    The results are shown in Fig.~\ref{fig:all_elements}(f), from where it is seen that Cr and Mn exhibit mainly ground states with in-plane magnetization, while Fe and Co have mostly ground states with out-of-plane magnetization.
    There is some uncertainty underlying the large anisotropies computed for Fe and Co on Ag and Au.
    We combined the magnetic force theorem with the frozen potential approximation, which is valid as long as the change in the valence charge $\Delta Q$ upon rotation of the spin moments from the $z$-axis to the $x$-axis remains small, and these assumptions are not well-satisfied by these systems.

	Next, we discuss the different contributions to the magnetism created by the presence of the magnetic adatoms on the non-magnetic surfaces.
    For simplicity, we set the magnetic moment of the adatoms to point out-of-plane.
    This preserves the $C_{\mathrm{3v}}$ symmetry of the (111) surface, which reduces the computational effort associated with the construction of the giant hemispherical clusters which are needed to obtain the inter-atomic orbital moments (see Ref.~\onlinecite{brinker2018} for details).
    The different contributions to the magnetic moment are shown in Fig.~\ref{fig:all_elements}, as well as Table~\ref{tab:SMM_OMM_all_elements}.
    Fig.~\ref{fig:all_elements}(a) shows that the spin moment of the adatoms follows the same trend on the different surfaces, reflecting the filling of the magnetic $d$-orbitals (cf.\ Fig.\ \ref{fig:app_ldos_all_elements}).
    Furthermore, the spin moment increases when surface element belongs to a group more to the right in the periodic table ($\mathrm{Rh}\rightarrow\mathrm{Pd}\rightarrow\mathrm{Ag}$ and $\mathrm{Ir}\rightarrow\mathrm{Pt}\rightarrow\mathrm{Au}$), which shows that the hybridization of the adatoms with the surfaces is decreasing.
    This can also be seen by the narrowing of the $d$-orbital peaks near the Fermi energy in the local density of states in Fig.\ \ref{fig:app_ldos_all_elements}.
    The orbital moments of the adatoms are shown in Fig.~\ref{fig:all_elements}(b).
    These are mostly quenched due to the lack of orbital exchange in the LSDA, and arise from the polarization of the $d$-orbitals at the Fermi energy due to the weak atomic spin-orbit coupling of the magnetic adatoms.
    The orbital moment is small for Cr and Mn, due to their small local density of states at the Fermi energy, and increases from Fe to Co, due to the increased spectral weight of the $d$-states at the Fermi energy.
    
    We now turn our attention to the induced surface magnetism.
    Fig.~\ref{fig:all_elements}(c-e) show the induced spin moments, the induced atomic orbital moments and the inter-atomic orbital moments, respectively.
	The induced spin moments reflect the spin polarizability of the substrates, which is highest for Pd and Pt and lowest for Ag and Au.
	Substrates with partially-filled $d$-bands at the Fermi level (Rh, Pd, Ir and Pt) show a positive correlation between the induced spin moment and the filling of the magnetic $d$-orbitals, while those with $sp$-like bands (Ag and Au) show a negative correlation.
	Since the induced atomic orbital moment is coupled to the induced spin moment via the local spin-orbit coupling of the surface atoms, it shows a similar trend.
	This is quite different in the case of the inter-atomic orbital moment --- it is large for Pt and small for the other surfaces.

    In Ref.~\onlinecite{brinker2018}, we reported on a correlation between the induced magnetic moments and local properties of the adatoms.
    The induced magnetic moments, $m^\text{ind}$, are approximately linear in both the spin magnetic moment of the adatom $m^\text{ad}_\text{s}$ and its relative spin polarization at the Fermi level $P^\text{ad}_\text{s} = \frac{\rho_\downarrow(E_\text{F}) - \rho_\uparrow(E_\text{F})}{\rho_\downarrow(E_\text{F}) + \rho_\uparrow(E_\text{F})}$, which is defined in terms of the spin-projected local density of states at the Fermi energy, $\rho_\downarrow(E_\text{F})$ and $\rho_\uparrow(E_\text{F})$ (see Fig.\ \ref{fig:app_ldos_all_elements}).
    This linearity is formulated in terms of effective susceptibilities,
\begin{align}\label{eq:suscfit}
\begin{split}
    m^\text{ind} &= \chi_m m^\text{ad}_\text{s} + \chi_P P^\text{ad}_\text{s} \\
    \text{or} \qquad
     \frac{m^\text{ind}}{m^\text{ad}_\text{s}} &= \chi_m  + \chi_P \frac{P^\text{ad}_\text{s}}{m^\text{ad}_\text{s}}\quad .
\end{split}
\end{align}
    Here we discover that this relation is valid not only for Pt(111) but also for the other surfaces.
    Only for the Ag surface the induced moments deviate noticeably from the linear trend (see Appendix \ref{app:susc_fit}).
    The fitted susceptibilities, $\chi_P$ and $\chi_m$, are collected in Table \ref{tab:susc_fits}.
    They are useful to classify and quantify the mechanisms which are responsible for the different contributions to the induced magnetic moment.
    The susceptibilities for the induced spin moments are highest for Pd, which is known to show the highest spin polarizability.
    The relative change between the 4\textit{d} elements Rh and Pd and the 5\textit{d} elements Ir and Pt is nearly identical.
    Au and Ag show a small magnetic response.
    The induced atomic orbital moments respond in nearly identical fashion for Pd and Pt, as well as Rh and Ir, respectively.
    The 5\textit{d} elements have a smaller spin polarizability but a larger spin-orbit coupling strength in comparison to the 4\textit{d} elements, and this balances out for the atomic orbital moment.
    The dependence of the inter-atomic orbital moment on the two main driving mechanisms of the substrate, spin polarizability and spin orbit coupling strength, is less clear, as it does not follow the previously explained trends.
    However, it is by far largest for the Pt surface, which combines a large spin polarizability with strong spin orbit coupling.
	
	\begin{figure}[tb]
		\includegraphics[width=\columnwidth]{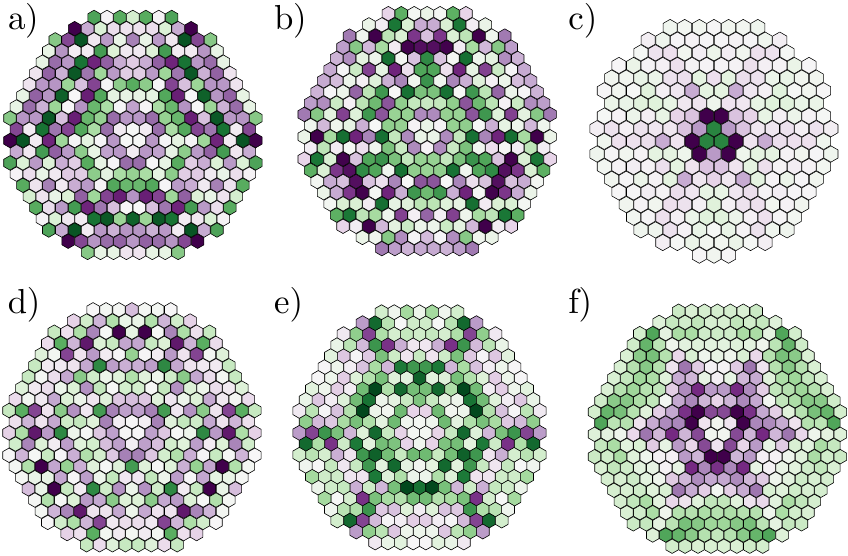}
		\caption{\label{fig:elements_combined_map} 	
		Inter-atomic orbital moment  of the atoms comprising the surface layer for an Co adatom deposited on the different substrates. 
		The values are scaled by the square of the distance to the adatom and the color scale differs for the different surfaces.
		Shown are the surfaces of
		a) Rh(111),
		b) Pd(111),
		c) Ag(111),
		d) Ir(111),
		e) Pt(111), and
		f) Au(111).
		}
	\end{figure}
    
    \begin{table}[bt]
		\begin{ruledtabular}
			\begin{tabular}{c|cc|cc|cc} 
			& \multicolumn{2}{c|}{$M_\text{s}$} & \multicolumn{2}{c|}{$M_\text{o}^\text{a}$} & \multicolumn{2}{c}{$M_\text{o}^\text{ia}$} \\ \hline 
			& $\chi_P [\mu_\text{B}]$ & $\chi_m$ & $\chi_P [\mu_\text{B}]$ & $\chi_m$ & $\chi_P [\mu_\text{B}]$ & $\chi_m$ \\ 
			Rh & $\phantom{+}1.13$ & $-0.12$ & $\phantom{+}0.067$ & $-0.014$ & $-0.024$ & $\phantom{+}0.005$ \\
			Pd & $\phantom{+}1.87$ & $\phantom{+}0.08$ & $\phantom{+}0.230$ & $-0.005$ & $\phantom{+}0.002$ & $-0.001$ \\
			Ag & $-0.18$ & $\phantom{+}0.07$ & $\phantom{+}0.005$ & $-0.004$ & $\phantom{+}0.010$ & $\phantom{+}0.004$ \\
			Ir & $\phantom{+}0.65$ & $-0.10$ & $\phantom{+}0.055$ & $-0.022$ & $-0.028$ & $-0.006$ \\
			Pt & $\phantom{+}0.93$ & $\phantom{+}0.03$ & $\phantom{+}0.204$ & $-0.016$ & $\phantom{+}0.186$ & $-0.020$ \\
			Au & $-0.10$ & $\phantom{+}0.07$ & $\phantom{+}0.019$ & $\phantom{+}0.005$ & $-0.044$ & $\phantom{+}0.005$ \\
			\end{tabular}
		\end{ruledtabular}
	\caption{
	    Susceptibilities fitted to Eq. \eqref{eq:suscfit} for the induced magnetic moments on the (111) facets of the different surfaces.
		}\label{tab:susc_fits}
	\end{table}
	
	Fig.~\ref{fig:elements_combined_map} shows the inter-atomic orbital moments of atoms forming the surface layer.
	The substrates with complex Fermi surfaces, or to be more precise complex constant energy contours (Rh, Pd, Ir and Pt), show an anisotropic distribution of the inter-atomic orbital moments, whereas Ag and especially Au with Rashba-like surface states are more isotropic.
    The spatial decay of the magnitude of the inter-atomic orbital moment depends on the substrate --- it is lowest for Rh and Pd and highest for Ag.
	
    Overall, we find that the inter-atomic orbital moment is mostly independent of the induced local spin moment, and is generated by a complex interplay of spin-orbit coupling, spin polarizability, interference effects and long-ranged (anisotropic) Friedel-like oscillations.

\subsection{Impact of the adatom-substrate distance on Pt(111)}
\label{sec:relaxation_dependence_Pt}
	\begin{figure}[t]

		\includegraphics[width=\columnwidth]{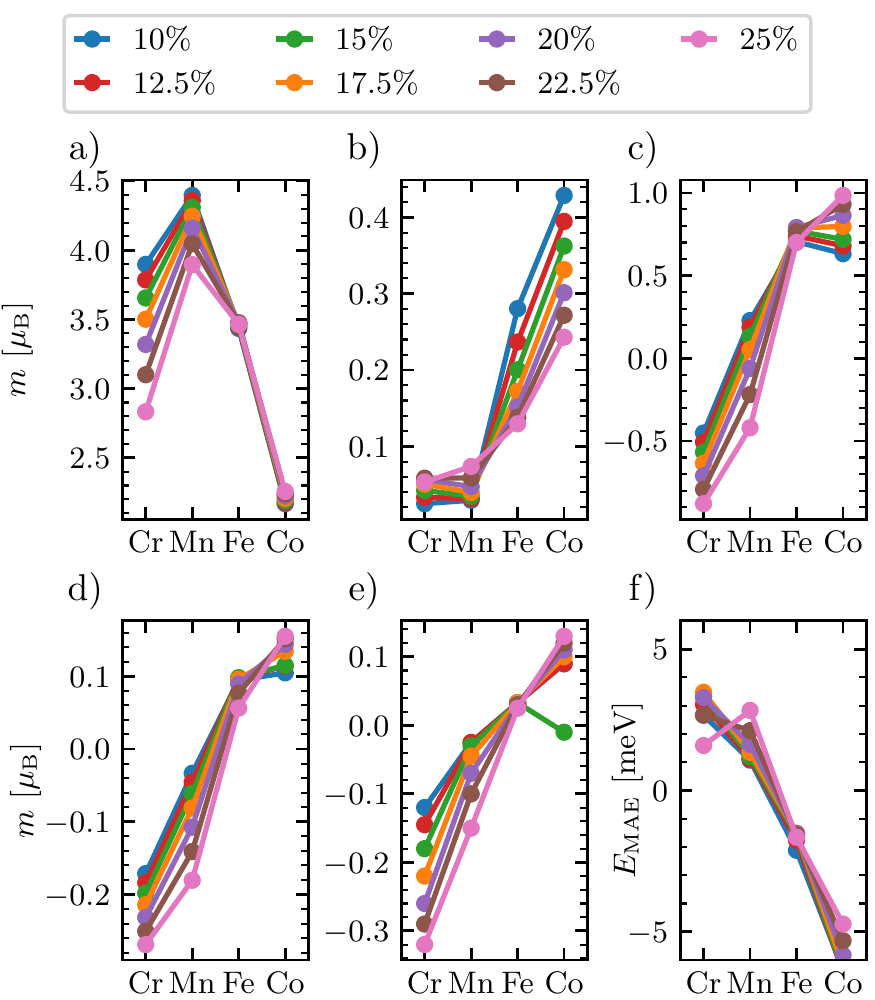}
		\caption{\label{fig:relaxation_dependence} 	
		Dependence of the magnetic contributions on the relaxation of the adatoms towards the surface layer of Pt(111). 
		a) Spin moment of the adatom. 
		b) Orbital moment of the adatom.
		c) Induced spin moments to the Pt.
		d) Induced atomic orbital moments to the Pt.
		e) Inter-atomic orbital moments in the Pt.
		All magnetic moments are in units of $\mu_\mathrm{B}$, as indicated on the left.
		f) Magnetic anisotropy energy.
		}
	\end{figure}
	We turn to quantifying the dependence of the different magnetic contributions on the relaxation of the adatoms towards the surface, taking the Pt(111) surface as reference and a range of relaxations between $10\%$ and $\SI{25}{\percent}$.
	Fig.~\ref{fig:relaxation_dependence}(a) shows that the Cr spin moment depends strongly on the relaxation, the Mn one has a weaker dependence, and nearly no dependence is found for Fe and Co.
	Considering the local density of states of the adatoms, Fig.\ \ref{fig:app_ldos_all_elements}, we note that the minority spin $d$-peak is empty for both Cr and Mn, whereas the majority spin $d$-peak is at the Fermi level for Cr and close to it for Mn.
	Increasing the hybridization by increasing the relaxation towards the surface leads to a broadening of the $d$-peaks in both spin channels (not shown), which results in a strongly lowered spin moment for Cr and a noticeable lowering of the spin moment for Mn.
	Fe and Co have a lower dependence on the relaxation, since the majority spin channels are strongly bound and lower in energy compared to Cr and Mn, which results in no noticeable effect of the broadening of the majority spin channel.

	Fig.~\ref{fig:relaxation_dependence}(b) plots the orbital moments of the adatoms, which originate mainly from the polarization of the $d$-orbitals at the Fermi energy due to the weak atomic spin-orbit coupling of the magnetic adatoms.
	Therefore the important quantity is the corresponding local density of states at the Fermi level.
	This is low for Cr and Mn, resulting in only small orbital moments, while Fe and Co have partially-filled minority-spin $d$-peaks, which lead to larger orbital moments for those adatoms.
	As relaxation towards the surface strongly modifies the hybridization of those partially-filled $d$-orbitals, the orbital moments are very sensitive to it.

	The induced magnetic moments, Fig.~\ref{fig:relaxation_dependence}(c-e), depend in two ways on the relaxation:
	Directly via the change of the hybridization of the adatom with the surface, and indirectly via the change of the spin moments of the adatom, which influences the strength of the perturbation being responsible for the induced magnetic moments.
	As a general trend, it is noticeable that the induced moments for Cr and Mn adatoms are more sensitive to the relaxation than Co and especially Fe, which is very insensitive to the relaxation.
	
	\subsection{Confinement effects on Pt(111)}
	\begin{figure}[t]
	    \includegraphics[width=\columnwidth]{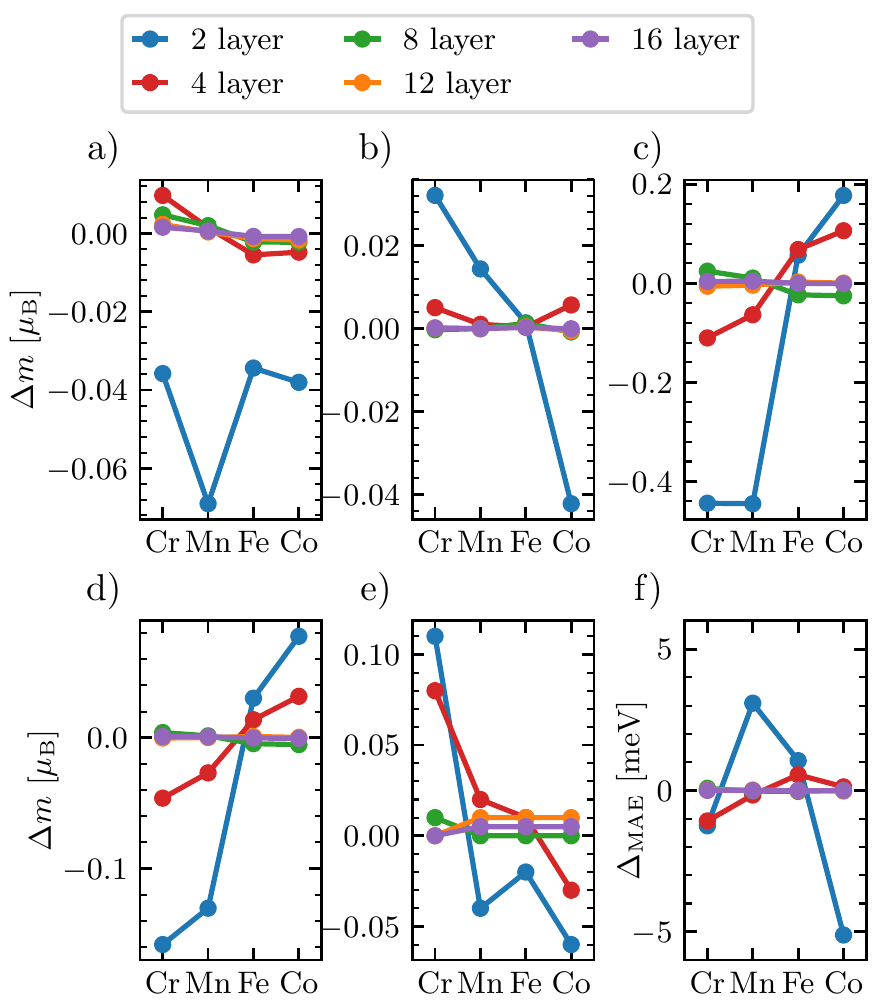}
		\caption{\label{fig:layer_dependence}
		Dependence of the magnetic contributions on the thickness of the Pt(111) slab for a fixed relaxation of $\SI{20}{\percent}$ toward the surface. 
		Shown are the deviations from the reference calculation containing 40 Pt layers.
		a) Spin moment of the adatom. 
		b) Orbital moment of the adatom.
		c) Induced spin moments to the Pt.
		d) Induced atomic orbital moments to the Pt.
		e) Inter-atomic orbital moments in the Pt.
		The magnetic moments are in units of $\mu_\text{B}$, as indicated on the left.
		f) Magnetic anisotropy energy of the adatom.
		}
	\end{figure}
	
	The dependence of the different magnetic contributions on the thickness of the Pt(111) slab is shown in Fig.~\ref{fig:layer_dependence} for different thicknesses ranging between 2 and 16 layers of Pt. This can be realized experimentally by depositing thin films of Pt on a different substrate.
	The values are compared to the reference calculation with our standard thickness of 40 Pt layers, which is assumed to be the closest to what is expected for a semi-infinite surface. This demonstrates that one can tune the various magnetic contributions to the surface magnetism via confinement effects. 
	We note that calculations employing 8 Pt layers are already very close to the results obtained for the thick slab, and some properties are already converged when computed with a slab of 4 Pt layers. 
	\begin{figure*}[t]
		\includegraphics[scale=1.0]{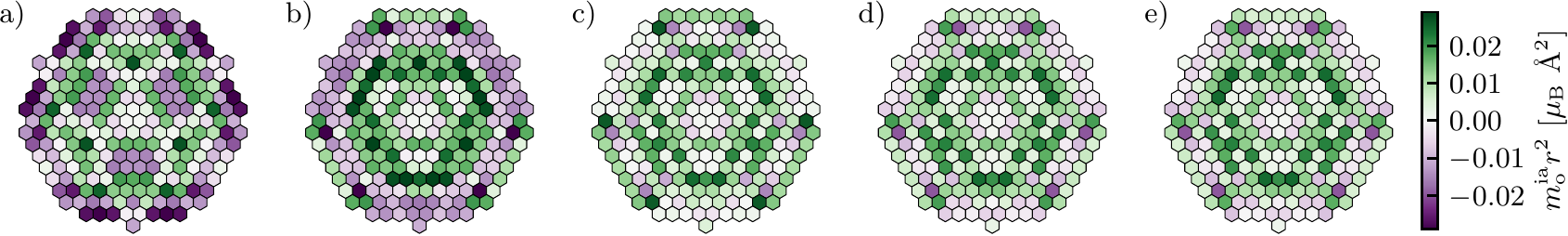}
		\caption{\label{fig:layer_map} 	
		Inter-atomic orbital moment  of the Pt atoms comprising the surface layer for an Co adatom deposited on a Pt(111) slab with different thicknesses containing
		a) 2 layer,
		b) 4 layer,
		c) 8 layer,
		d) 12 layer, and
		e) 16 layer.
		}
	\end{figure*}
	
    The dependence of the inter-atomic orbital moments generated by a Co adatom on the atoms comprising the Pt surface layer on the different thicknesses is depicted in Fig.~\ref{fig:layer_map}.
    Fig.~\ref{fig:elements_combined_map}(e) is the reference calculation.
    In comparison to it, the system with two layers of Pt (see Fig.~\ref{fig:layer_map}(a)) shows a changed oscillation pattern.
    Due to interference effects originating from the scattering of the lower boundary of the slab, the inter-atomic moment becomes more anisotropic and its oscillation wavelength changes.
    Increasing the slab thickness to 4 layers (Fig.~\ref{fig:layer_map}(b)) eliminates the most prominent interference effects.
    The pattern is essentially converged when the thickness is set to 8 layers (Fig.~\ref{fig:layer_map}(c)), and further increases produce very minor changs to the picture, Fig.~\ref{fig:layer_map}(d-e).

	\subsection{Orientation dependence of the magnetic moments of the adatoms on the Pt(111) surface}
	\begin{table}[tb]
		\begin{ruledtabular}
			\begin{tabular}{ccccccc} 
			\multicolumn{2}{l}{Adatom}	&	$ m_\text{s}^\text{ad} [\mu_\text{B}]$ & $m_\text{o}^\text{ad} [\mu_\text{B}]$ & $ M_\text{s} [\mu_\text{B}]$ & $ M_\text{o}^\text{a} [\mu_\text{B}]$ & $M_\text{o}^\text{ia} [\mu_\text{B}]$ \\ \hline
			\multirow{2}{*}{Cr} & $z$ & 3.32 & 0.06 & -0.70 & -0.23 & -0.25 \\
			                    & $x$ & 3.32 & 0.17 & -0.62 & -0.27 & \phantom{-}0.03 \\ \hline 
			\multirow{2}{*}{Mn} & $z$ & 4.16 & 0.05 & -0.05 & -0.11 & -0.07 \\
                                & $x$ & 4.16 & 0.09 & -0.03 & -0.11 & \phantom{-}0.06 \\ \hline
            \multirow{2}{*}{Fe} & $z$ & 3.48 & 0.15 & \phantom{-}0.88	& \phantom{-}0.08 & \phantom{-}0.03 \\
                                & $x$ & 3.48 & 0.17 & \phantom{-}0.81 & \phantom{-}0.13 & \phantom{-}0.05 \\ \hline
            \multirow{2}{*}{Co} & $z$ & 2.22 & 0.30 & \phantom{-}0.86	& \phantom{-}0.14 &\phantom{-}0.12 \\
                                & $x$ & 2.22 & 0.18 & \phantom{-}0.84 & \phantom{-}0.17 & \phantom{-}0.04 
			\end{tabular}
		\end{ruledtabular}
	\caption{
		Dependence of the different contributions to the magnetic moments on the orientation of the spin moment of the adatoms on Pt(111).
		We list the spin moment $m_\text{s}^\text{ad}$ and orbital moment $m_\text{o}^\text{ad}$ of the adatom, and the induced spin moment $M_\text{s}$, induced atomic orbital moment $M_\text{o}^\text{a}$ and inter-atomic orbital moment $M_\text{o}^\text{ia}$.
		The positive or negative sign reflects whether the magnetic moment is parallel or antiparallel to the chosen orientation of the spin moment of the adatom.
		The spin moment and the atomic orbital moment are computed from a connected cluster including 169 substrate atoms, whereas the inter-atomic orbital moment is computed from a giant hemispherical cluster containing 2685 substrate atoms.
		}\label{tab:SMM_OMM_in-plane}
	\end{table}
	The induced magnetism of the surface depends on the orientation of the spin moment of the adatoms.
	Table~\ref{tab:SMM_OMM_in-plane} lists the different contributions to the magnetic moment of the adatoms deposited on the Pt(111) surface, when the spin magnetic moment is aligned either out-of-plane ($z$-axis) or in-plane along the $x$-axis. 
	The magnitude of the spin moments of the adatoms is independent of their orientation, as they are generated by strong isotropic intra-atomic exchange interactions, which are very weakly influenced by the atomic spin-orbit coupling.
	The induced spin moments on the surface have a stronger orientation dependence, likely due to the much stronger spin-orbit coupling of Pt.
    
    The orbital moments are much more anisotropic than the spin moments.
    The orbital moments of the adatoms tend to be largest when the spin moment is along the axis that minimizes the magnetic anisotropy energy\cite{Bruno1989}, in-plane for Cr and Mn and out-of-plane for Co.
    Fe is an exception, with a weak anisotropy of its orbital moment.
    The induced atomic orbital moments on the surface are also quite anisotropic, and are parallel to the induced spin moments, showing that they originate from the strong spin-orbit coupling of Pt.

	Compared to the other magnetic contributions, the inter-atomic orbital moment changes drastically for the in-plane configuration compared to the out-of-plane configuration.
	This can be intuitively understood from picturing the inter-atomic orbital moment as being generated by the currents that swirl around the orientation of the spin moment of the adatom and pass through the surface atoms.
	When the spin moment of the adatom is out-of-plane, the currents can form large loops in the $xy$-plane.
	If the sense of the swirl does not change too much with the distance to the adatom (weak Friedel-like oscillations), this adds up to a large net inter-atomic orbital moment, as seen for Cr and Co, which is parallel to the induced spin moment on the surface.
	When the spin moment of the adatom is oriented along the $x$-axis, the currents should now swirl in the $yz$-plane, which is cut in half by the presence of the surface.
	Furthermore, the ground-state currents are divergence-free and so the current loops have to close on the surface, making the Pt surface atoms contribute less for this in-plane configuration as compared to the out-of-plane configuration.
	Notably, the inter-atomic orbital moment is now parallel to the spin moment of the adatom.
    
    \subsection{Current-induced magnetic fields for a Co adatom deposited on Pt(111)}
	\begin{figure}[tb]
		\includegraphics[width=\columnwidth]{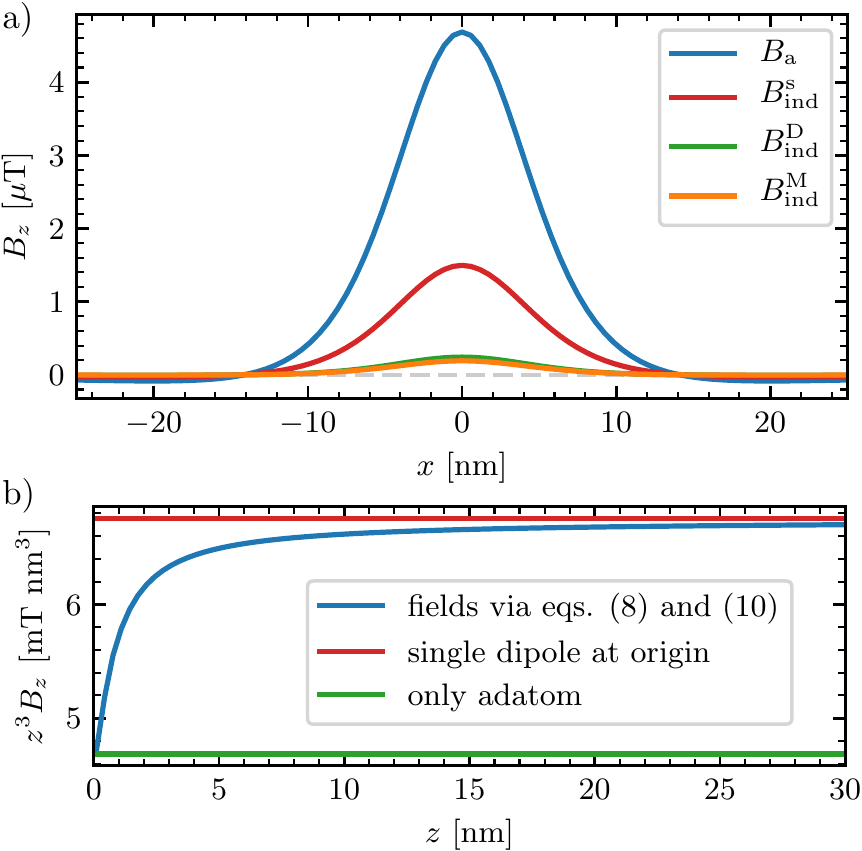}
		\caption{\label{fig:stray_field} 	
		Magnetic stray fields above a Co adatom deposited on the Pt(111) surface.
		a) Horizontal scan of the magnetic stray fields at a height $z = \SI{10}{\nano\meter}$ above the surface.
		$B_\mathrm{a}$ is the stray field generated by the total magnetic moment (spin+orbital) of the adatom, $B^\mathrm{s}_\mathrm{ind}$ is the stray field generated by the induced spin moments on the surface, and $B^\mathrm{M}_\mathrm{ind}$ and $B^\mathrm{D}_\mathrm{ind}$ are the stray fields created by the monopole and dipole contributions from the induced surface ground-state charge currents.
		b) Magnetic stray field vertically above the adatom.
		The magnetic stray field taking all the spatial dependencies via eqs.~\eqref{eq:stray_field_dipole_approx} and \eqref{current_induced_bfield} into account (blue curve) is compared to the stray field of a single dipole corresponding to the total magnetic moment (red curve) and the stray field induced only by the adatom (green curve).
		}
	\end{figure}
	A possible way of investigating the surface magnetism of adatoms and other magnetic nanostructures on surfaces is by detecting the magnetic stray field that they generate, for instance by scanning NV-center microscopy.
    In the following, we apply the formalism developed to account for the current-induced magnetic fields together with the usual dipolar contribution of the spin moments (Sec.~\ref{sec:current_induced_fields}) to the case of a Co adatom deposited on the Pt(111) surface.
    The magnetic stray field originates from three contributions: 
    the spin moments and the atomic orbital moments in the dipole approximation and the inter-atomic orbital moments via the net-currents in the monopole approximation.

    If the stray field is probed at some distance from the surface, it should approach the dipolar form as if generated by a single magnetic moment.
    Fig.~\ref{fig:stray_field}a shows the different contributions to the stray field generated by a Co adatom at a height of \SI{10}{\nano\meter} above the Pt(111) surface.
    Despite the very different spatial spread of the different contributions to the magnetic moment --- the spin and orbital magnetic moments localized in the adatom, the induced spin and atomic orbital moments on the Pt atoms, which spread out away from the adatom, and the induced inter-atomic orbital moments, which extend even farther ---, the corresponding contributions to the stray field have very similar shapes, differing mostly in their amplitude.

    A feasible experiment would be to try to quantify the total induced magnetic moment on the surface, by carefully measuring the distance dependence of the stray field vertically above the magnetic adatom.
    The concept is illustrated in Fig.~\ref{fig:stray_field}b, where the computed stray field is compared with the pure dipolar field generated either by a magnetic dipole moment equal to the total magnetic moment and centered at the adatom, and with the dipolar field generated by the magnetic moment of the adatom alone.
    The computed stray field approaches the former in the large-separation limit, and the latter in the short-separation limit.
    The deviation from those two limits contains information about the magnitude of the induced magnetic moment and its spatial distribution around the magnetic adatom, and the corresponding measurements could be directly compared with our computed curves.

	\subsection{Compact trimers deposited on Pt(111)}
	\begin{table}[tb]
		\begin{ruledtabular}
			\begin{tabular}{cccccc} 
				Trimer &	$ m_\text{s}^\text{tri} [\mu_\text{B}]$ & $m_\text{o}^\text{tri} [\mu_\text{B}]$ & $ M_\text{s} [\mu_\text{B}]$ & $ M_\text{o}^\text{a} [\mu_\text{B}]$ & $M_\text{o}^\text{ia} [\mu_\text{B}]$ \\ \hline 
Cr   &   3.20   &         0.04     &      -1.03    &       -0.53       &    0.05 \\ 
Mn   &   4.07   &         0.05     &      \phantom{-}0.40     &       -0.24       &    0.12 \\ 
Fe   &   3.32   &         0.12     &      \phantom{-}1.63     &       \phantom{-}0.14        &    0.06 \\ 
Co   &   2.19   &         0.20     &      \phantom{-}2.12     &       \phantom{-}0.32        &    0.21 
			\end{tabular}
		\end{ruledtabular}
	\caption{
		Different contributions to the magnetic moment created by ferromagnetic fcc-top-stacked trimers deposited on the Pt(111) surface.
		Spin moment $m_\text{s}^\text{tri}$ and orbital moment $m_\text{o}^\text{tri}$ of the trimer per atom and induced spin moment $M_\text{s}$, induced atomic orbital moment $M_\text{o}^\text{a}$ and inter-atomic orbital moment $M_\text{o}^\text{ia}$ for Cr, Mn, Fe and Co trimers.
		The spin moment and the atomic orbital moment are taken from a connected cluster including 149 substrate atoms, whereas the inter-atomic orbital moment is taken from a giant hemispherical cluster containing 2683 substrate atoms.
		}\label{tab:SMM_OMM_trimer}
	\end{table}
	In this section, we investigate the influence of the size of the nanostructure by studying fcc-top-stacked trimers deposited on the Pt(111) surface (see Fig.~\ref{fig:adatom_trimer_illustration}b).
	The different magnetic contributions for Cr, Mn, Fe and Co trimers are shown in Table~\ref{tab:SMM_OMM_trimer}.
	Note that we consider here a collinear out-of-plane configuration of the magnetic moments, which is a good assumption for Fe and especially Co due to the strong ferromagnetic interaction between the trimer atoms resulting in only small noncollinearities.
	Cr and Mn exhibit strong antiferromagnetic interactions in the trimer resulting in a highly frustrated magnetic ground state. 
	Comparing the magnetic moments of each trimer atom to the magnetic moments of a single adatom (see Table~\ref{tab:SMM_OMM_in-plane}), there are only minor differences for the spin moment of each atom, whereas the orbital moment is reduced by $\SI{30}{\percent}$ for Fe and Co, due to the additional hybridization with the neighboring magnetic atoms comprising the trimer.
	The induced magnetic moments are generally enhanced due to the presence of multiple magnetic atoms, but are not simply three times larger than the ones found for the isolated adatoms.
	For example, the induced spin moment of the Co trimer is approximately $2.5$ times larger, the induced atomic orbital moment is $2.3$ times larger, and the inter-atomic orbital moment is $1.8$ times larger than the corresponding adatom values.
	The modification of the electronic structure of the magnetic $d$-orbitals by bringing the adatoms together to form a trimer changes the effective coupling to the surface.
	Furthermore, and as seen in Fig.~\ref{fig:elements_combined_map} for the case of the inter-atomic orbital moment, each magnetic atom tends to form a complex Friedel-like oscillation pattern of induced moments, which if superimposed by having three point sources instead of one can also lead to destructive interference-like effects, thus suppressing the value of the net induced moments.

	Noteworthy, the formation of a large nanostructure can lead to a completly different response of the Pt surface atoms with respect to the induced net currents and the related inter-atomic orbital moment, which can be seen for Cr and Mn.
	Both atoms (but especially Cr) showed a strong antiferromagnetic behaviour for the inter-atomic orbital moments with respect to the spin moments in the single adatom case.
	However, in a trimer the surface responded in a ferromagnetic fashion for both atom species.	
	
\section{Conclusions}
We presented a comprehensive analysis of the induced magnetism of magnetic 3$d$ adatoms deposited on several non-magnetic surfaces, paying special attention to place the recently discovered inter-atomic orbital magnetic moments in the context of the other contributions to the induced surface magnetism.
The magnetic moments of the adatoms and the induced ones were related to the electronic structure of the adatoms and of the different surfaces, and the trends explained through effective magnetic susceptibilities.
The Pt(111) surface was confirmed as the ideal surface for a large inter-atomic orbital moment to emerge, due to its combination of high spin polarizability and significant spin-orbit coupling.
The dependence of the results on the computational approximations was investigated for this surface, as well as how the induced magnetism scales if the size of the nanostructure is increased, by considering Cr, Mn, Fe and Co trimers.

The spin and orbital magnetic moments of the adatoms could be explained by the progressive filling of the magnetic $d$-orbitals and by the relative strength of their hybridization with each considered surface.
The induced magnetic moments on the surface could be correlated to two key properties: the spin polarization at the Fermi level and the local magnetic moment of the adatom.
These two properties were used to correspondingly define two effective magnetic susceptibilities, that could successfully explain the trends in the computed results.
These effective susceptibilities define at the same time material-specific parameters for each considered surface and shed light on the physical origin of the induced magnetism.

We have also explored the dependence of magnetic properties on computational approximations or assumptions, such as the structural relaxation of the adatoms towards the Pt(111) surface, or the thickness of the slab.
The spin moments of Cr and Mn are quite sensitive to the distance between the adatom and the surface, while this is not the case for the Fe and Co adatoms.
The converse is true concerning the orbital magnetic moments of the adatoms.
This can be traced to a combination of the progressive filling of the magnetic d-orbitals when going from Cr to Fe with an increased hybridization with the surface as the adatom approaches it.
Comparison with experiment shows that the orbital moment of the adatom (but not its spin moment) tends to be underestimated by standard exchange-correlation functionals~\cite{gambardella_giant_2003,Blonski2010}, which can be partially compensated for by larger distances between the adatom and the surface~\cite{Lazarovits2002,Sipr2010,Sipr2016}.
The induced magnetic moments on the surface and the magnetic anisotropy energy all evolve with the distance between the adatom and the surface, with the curious exception of the Fe adatom.
The thickness of the Pt(111) turns out not to be very important, with converged results obtained for a thickness of 8 Pt layers, and even a thickness of 4 layers leading to reasonable values for most quantities. 
This also indicates that confinement effects can be used to tune the various induced magnetic moments by controlling the thickness of the films on which the nanostructures are deposited.
The orbital magnetic moments of the adatoms and those induced on the surface are also quite anisotropic, and so must be computed for at least two orthogonal directions for their variability to be ascertained.

Our computed stray fields show that it is experimentally difficult to disentangle contributions of different origin.
Having in mind NV-center microscopy, one key parameter is the distance between the surface and the NV-center, which experiences the stray field at its position.
Lateral scans over a magnetic adatom probably cannot resolve subtle changes in the magnetic field profile due to the different spatial distribution of the magnetic moment of the adatom vs. the induced ones on the surface.
If the NV-center can somehow be sufficiently approached to the magnetic adatom, say down to a separation of $d \sim \SI{5}{\nano\meter}$, our calculations show that it should be possible to detect a variation in the standard $1/d^3$ decay of the stray field into vacuum, that could be related to the magnitude of the induced magnetic moments.
Although we expect these conditions to be very challenging to meet experimentally, it does prove that NV-center magnetometry can in principle distinguish them.

\begin{acknowledgments}
This work was supported by the European Research Council (ERC) under the European Union's Horizon 2020 research and innovation program (ERC-consolidator grant 681405 -- DYNASORE). We gratefully acknowledge the computing time granted by the JARA-HPC Vergabegremium and VSR commission on the supercomputer JURECA at Forschungszentrum J\"ulich.
\end{acknowledgments}

\appendix

\section{Data tables} \label{app:tables}
Here we present the tables containing the data that is plotted in Figs.~\ref{fig:relaxations_adatoms} and \ref{fig:all_elements}.
    \begin{figure*}[!tbh]
		\includegraphics[scale=0.9]{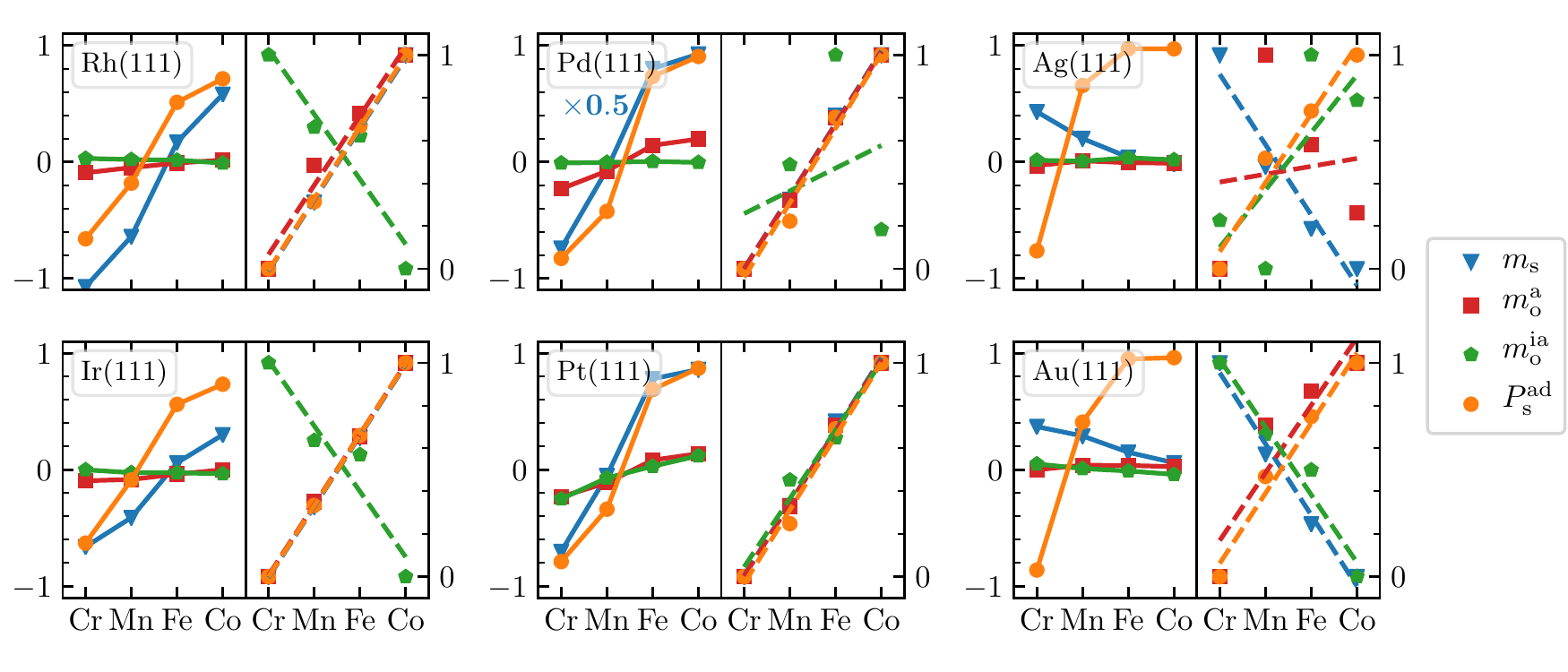}
		\caption{\label{fig:app_numerology}	
		Relations between the induced surface magnetic moments and the magnetic properties of the adatoms deposited on the different surfaces.
        Left panels: Spin $m_\text{s}$, atomic $m_{\text{o}}^{\text{a}}$ and inter-atomic $m_{\text{o}}^{\text{ia}}$ orbital moments (in $\mu_{\text{B}}$), and relative spin polarization at the Fermi energy $P^\text{ad}_\text{s}$ of each adatom.
        Right panels: Same as in the left panels, but with all quantities divided by the corresponding spin moment of each adatom, $m_\text{s}^\text{ad}$ (see Eq.~\eqref{eq:suscfit}).
        The dashed lines are linear fits to the data.
        The data is normalized (0 to 1) to show the linear behaviour for the different induced moments in the same plot.
		}
	\end{figure*}
	\begin{table}[!tb]
		\begin{ruledtabular}
			\begin{tabular}{lcccccc} 
				& Rh & Pd & Ag & Ir & Pt & Au \\
				$a_0^\text{exp}$ [\AA] & $3.793 $ & $3.876 $ & $4.063 $ & $3.832 $ & $3.913 $ & $4.061 $ \\
				$a_0$ [\AA] & $3.785$ & $3.881$ & $4.064$ & $3.839$ & $3.914$ & $4.084$ \\
				$\Delta_\text{surface}$ & $\SI{-1.9}{\percent}$ & $\SI{+0.1}{\percent}$ & $\SI{-1.2}{\percent}$ & $\SI{-1.8}{\percent}$ & $\SI{0.3}{\percent}$ & $\SI{-0.1}{\percent}$ \\
				$\Delta_\text{Cr}$& $\SI{13.2}{\percent}$ & $\SI{19.1}{\percent}$ & $\SI{12.1}{\percent}$ & $\SI{11.4}{\percent}$ & $\SI{19.4}{\percent}$ & $\SI{17.3}{\percent}$ \\
				$\Delta_\text{Mn}$& $\SI{14.6}{\percent}$ & $\SI{17.9}{\percent}$ & $\SI{15.8}{\percent}$ & $\SI{11.7}{\percent}$ & $\SI{17.9}{\percent}$ & $\SI{21.1}{\percent}$ \\
				$\Delta_\text{Fe}$& $\SI{18.7}{\percent}$ & $\SI{26.2}{\percent}$ & $\SI{19.8}{\percent}$ & $\SI{17.9}{\percent}$ & $\SI{25.9}{\percent}$ & $\SI{27.7}{\percent}$ \\
				$\Delta_\text{Co}$& $\SI{20.8}{\percent}$ & $\SI{26.1}{\percent}$ & $\SI{21.4}{\percent}$ & $\SI{20.6}{\percent}$ & $\SI{27.5}{\percent}$ & $\SI{27.0}{\percent}$ \\
				$m_\text{Cr} [\mu_\text{B}]$ & $3.15$ & $3.32$ & $3.86$ & $3.07$ & $3.11$ & $3.77$\\
				$m_\text{Mn} [\mu_\text{B}]$ & $3.65$ & $3.99$ & $4.17$ & $3.70$ & $3.88$ & $4.14$\\
				$m_\text{Fe} [\mu_\text{B}]$ & $3.20$ & $3.34$ & $3.16$ & $3.08$ & $3.28$ & $3.22$\\
				$m_\text{Co} [\mu_\text{B}]$ & $2.09$ & $2.27$ & $2.02$ & $1.93$ & $2.17$ & $2.00$ 
			\end{tabular}
		\end{ruledtabular}
	\caption{
		Structural relaxations and adatom magnetic moments obtained from QE calculations.
		The first two data rows compare the calculated bulk lattice constants to the experimental ones with zero-point correction, taken from Table III of Ref.~\onlinecite{Hao2012}.
		Next we report on the vertical relaxation of the first surface layer of the fcc(111)-surface without the adatoms $\Delta_\text{surface}$, the vertical relaxations of the different adatoms $\Delta_\text{Cr/Mn/Fe/Co}$, and their magnetic moments $m_\text{Cr/Mn/Fe/Co}$.
		The relaxations $\Delta$ are defined with respect to the bulk inter-layer distance, $d = (1 - \Delta)\,a_0/\sqrt{3}$, so a positive sign means a relaxation towards the surface.
		}\label{tab:relaxations}
	\end{table}

	\begin{table}[!tb]
		\begin{ruledtabular}
			\begin{tabular}{c|cccccc} 
			Surface	&	Atom	&	$ m_\text{s}^\text{ad} [\mu_\text{B}]$ & $m_\text{o}^\text{ad} [\mu_\text{B}]$ & $ M_\text{s} [\mu_\text{B}]$ & $ M_\text{o}^\text{a} [\mu_\text{B}]$ & $M_\text{o}^\text{ia} [\mu_\text{B}]$ \\ \hline \hline
			\multirow{ 4}{*}{Rh(111)}	                & Cr 	&	2.87	&	0.014	&	-1.07	        & -0.093	&  \phantom{-}0.03\\
				 										& Mn 	&	3.81	&	0.023	&	-0.64           & -0.049	&  \phantom{-}0.02 \\
				 										& Fe 	&	3.24	&	0.111	&	\phantom{-}0.17	& \phantom{-}0.010		& \phantom{-}0.02 \\
				 										& Co 	&	2.03	&	0.190	&	\phantom{-}0.58	& \phantom{-}0.016		& -0.01\\ \hline
			\multirow{ 4}{*}{Pd(111)} 	                & Cr    &	3.46 	&	0.009	&	-1.48 	        & -0.231		& -0.01\\
				 										& Mn 	&	4.28 	&	0.021	&	-0.13 	        & -0.078		& -0.01\\
				 										& Fe 	&	3.55 	&	0.131	&	\phantom{-}1.60 & \phantom{-}0.142		& \phantom{-}0.00 \\
				 										& Co 	&	2.32 	&	0.357	&	\phantom{-}1.85 & \phantom{-}0.196	& -0.01 \\ \hline
			\multirow{ 4}{*}{Ag(111)} 	                & Cr 	&	4.26	&	0.018	&	\phantom{-}0.43	& -0.032	&  \phantom{-}0.01\\
				 										& Mn 	&	4.60	&	0.009	&	\phantom{-}0.20	& \phantom{-}0.009	& \phantom{-}0.01\\
				 										& Fe 	&	3.47	&	0.550	&	\phantom{-}0.04	& -0.007	& \phantom{-}0.04\\
				 										& Co 	&	2.19	&	0.668	&	-0.02	        & -0.011	& \phantom{-}0.02\\ \hline
			\multirow{ 4}{*}{Ir(111)} 	                & Cr 	&	2.71	&	0.010	&	-0.66	        & 	-0.094	& \phantom{-}0.00\\
				 										& Mn 	&	3.68	&	0.050	&	-0.41	        & 	-0.083	& -0.03\\
				 										& Fe 	&	3.11	&	0.136	&	\phantom{-}0.06	& 	-0.037	& -0.03\\
				 										& Co 	&	1.87	&	0.173	&	\phantom{-}0.30	& 	\phantom{-}0.000	& -0.04\\ \hline
			\multirow{ 4}{*}{Pt(111)} 	                & Cr 	&	3.32	&	0.057	&	-0.70	        & 	-0.232	& -0.25 \\
														& Mn 	&	4.16	&	0.047	&	-0.05	        & 	-0.110	& -0.07\\
				 										& Fe 	&	3.48	&	0.152	&	\phantom{-}0.88	& \phantom{-}0.083		& \phantom{-}0.03\\
				 										& Co 	&	2.22	&	0.301	&	\phantom{-}0.86	& 	\phantom{-}0.138	&\phantom{-}0.12\\ \hline
			\multirow{ 4}{*}{Au(111)} 	                & Cr 	&	4.05	&	0.004	&   \phantom{-}0.37	& 	\phantom{-}0.000	& \phantom{-}0.05 \\
				 										& Mn 	&	4.54	&	0.013	&	\phantom{-}0.29	& 	\phantom{-}0.039	& \phantom{-}0.01\\
				 										& Fe 	&	3.50	&	0.408	&	\phantom{-}0.15	& 	\phantom{-}0.037	& -0.01\\
				 										& Co 	&	2.22	&	0.434	&	\phantom{-}0.06	& 	\phantom{-}0.027	& -0.04\\
			\end{tabular}
		\end{ruledtabular}
	\caption{
		Ground state properties of 3d magnetic adatoms on several (111) surfaces.
		Spin moment $m_\text{s}^\text{ad}$ and orbital moment $m_\text{o}^\text{ad}$ of the adatom and total induced spin moment $M_\text{s}$, total induced atomic orbital moment $M_\text{o}^\text{a}$ and total inter-atomic orbital moment $M_\text{o}^\text{ia}$ for Cr, Mn, Fe and Co adatoms deposited on the (111) surface of Rh, Pd, Ag, Ir, Pt and Au.
		}\label{tab:SMM_OMM_all_elements}
	\end{table}

\section{Fitted susceptibilities} \label{app:susc_fit}
    The linear fits for the susceptibilities discussed in Eq.~\eqref{eq:suscfit} are shown in Fig.~\ref{fig:app_numerology} for all the different surfaces.
    The 5\textit{d} surfaces with large spin-orbit coupling show all qualitatively good linear fits.
    For the 4\textit{d} surfaces we find a good agreement for Rh and Pd with the exception of the inter-atomic orbital contribution for Pd, but very bad agreement for Ag.
    The induced spin moment is still well described, but both induced orbital contributions do not follow any linear trend.


\bibliography{Lib_OMM.bib}

\end{document}